\documentclass[11pt]{article}
\usepackage[margin=25mm]{geometry}

\makeatletter   
\newcommand{\FloatBarrier}{\par\begingroup\let\@elt\relax\edef\@tempa{\@botlist\@deferlist\@dbldeferlist}%
  \ifx\@tempa\@empty\endgroup\else\endgroup\clearpage\fi}
\makeatother
\usepackage{amsmath,amssymb,amsthm,bm}
\usepackage{booktabs,tabularx}
\usepackage{graphicx}
\usepackage[hidelinks]{hyperref}
\usepackage{xcolor}
\usepackage[numbers,sort&compress]{natbib}
\newcommand{\Iref}{I_{\mathrm{ref}}}
\newcommand{\x}{\bm{x}}
\newcommand{\w}{\bm{w}}
\newcommand{\uu}{\bm{u}}

\newcommand{\tbd}[1]{\textcolor{red}{[#1]}}
\newcommand{\inputtable}[1]{\IfFileExists{#1}{\input{#1}}{\tbd{table #1 pending}}}
\newenvironment{tablepair}{\begin{table}[!htbp]\centering\begingroup
  \renewenvironment{table}[1][]{}{\par}}{\endgroup\end{table}}
\makeatletter
\g@addto@macro\@floatboxreset{\linespread{1}\selectfont}  
\long\def\@makecaption#1#2{\vskip 3pt{\small\textbf{#1.} #2\par}\vskip 3pt}
\makeatother

\title{Hyperelasticity as an efficient regularizer for 2D--3D thoracic motion models}
\author{Nicholas Hindley \\ \small Image X Institute, School of Health Sciences, Faculty of Medicine and Health,\\
\small University of Sydney, Sydney, NSW, Australia \\ \small \texttt{nicholas.hindley@sydney.edu.au}}
\date{}

\begin{document}
\maketitle

\begin{abstract}
Patient-specific motion models enable real-time volumetric imaging of the thorax from a single x-ray projection, but most existing approaches require hours of training. The \emph{VM-DS} framework trains in minutes, yet its image-space self-consistency regularizer accounts for 40\% of the motion-model fitting time and does not prevent the fitted fields from folding. In this paper, we explore a range of regularization approaches including scaling-and-squaring, bending energy, diffusion, first-order inverse consistency, isotropic total variation, a negative-Jacobian hinge, $|\log\det J|$, a control-point injectivity bound and hyperelastic regularization. Hyperparameters were selected on two development cases from the public  DIR-Lab 4D dataset under rules fixed in advance, and the regularizers were tested on eight held-out cases. At a fixed number of training steps, the hyperelastic term trained in 5.6 against 8.4~min and gave a lower held-out TRE (2.5~mm). With a quarter of the motion-model budget it matched the baseline (2.7~mm), and at every budget from 2 to 17~min its mean TRE was lower by 0.3--0.5~mm. Hyperelastic fields had no folded voxel in any held-out view, whereas baseline fields folded in 83\% of views, with image similarity unchanged. A hyperelastic volume penalty can therefore replace image-space self-consistency, matching accuracy with 3--5 minutes of training and removing folding, bringing on-the-day motion models closer to routine use in image-guided interventions. Code will be available at \url{https://github.com/Image-X-Institute/Voxelmap/tree/hyper}
\end{abstract}

\section{Introduction}

Respiratory motion is a central obstacle to accurate image guidance in the thorax. Lung cancer
remains the leading cause of cancer death worldwide \cite{Bray2024}, and both its treatment and its
diagnosis depend on knowing where a moving target is. In radiotherapy, motion is managed with
margins, gating or tracking \cite{keall2006tg76}, and tracking the target in real time allows the
beam to follow it \cite{keall2021tg264}, but on standard systems only 2D x-ray projections are
acquired during delivery. In bronchoscopic biopsy, robotic navigation now reaches the target in
nearly every case, yet the diagnostic yield remains between 60 and 75\%
\cite{Murgu2025TARGET,Ost2026PRECIsE,Kops2023,Nadig2023}, and the deficit has been attributed to
tool positioning \cite{Maturu2025}. Intraprocedural cone-beam CT (CBCT) improves tool-in-lesion
rates and yield substantially \cite{Bashour2024,Husta2024}, but each spin is a static snapshot of an
anatomy that is continually moving as the patient is ventilated, and only two are acquired in a
procedure \cite{Maturu2025,Revelo2025}. Real-time volumetric imaging could address the tracking and sampling issues in both the radiotherapy and interventional settings.

Thus far, real-time volumetric imaging has been developed exclusively within the field of radiotherapy, where training data from treatment planning are available days or weeks before delivery. Early models used this data to extract the principal components of deformation fields \cite{li2010}, and deep learning has since recovered volumes, organ surfaces or deformation fields from single projections
\cite{Shen2019,Lei2020TransNet,Wang2019DeepOrganNet,hindley2023,hindley2025voxelmap}. \emph{DREME}
\cite{shao2025dreme} learns the reference image and the motion model jointly from a single CBCT. However, these methods require hours of training.

The \emph{VM-DS} \cite{vmds} optimization framework trains 2D--3D models in minutes using a dual-scan regime. A breath-hold CBCT, already acquired in the standard workflow, serves as the reference image, and a second, motion-inclusive CBCT arc supplies the training projections. The model is fit by alternating least squares, with Gauss--Newton updates of the per-projection scores and Adam updates of a shared B-spline basis, and then amortized by a projection encoder. \emph{VM-DS} caps training time within a 10-minute window. As the motion basis is learned from projections, the fit requires a regularizer. \emph{VM-DS} uses a self-consistency term, which warps the deformed image back with an inverse field and compares the result with
the reference. However, the term has two drawbacks. First, it is slow, since it requires an inverse field and a
second warp for every projection, and accounts for 40\% of the motion-model training time.
Second, it does not prevent the fitted fields from folding, which matters
for dose accumulation, contour propagation and any other use of the inverse map.

Regularization has been studied for decades in deformable 3D--3D registration. Examples include the
bending energy of free-form deformations \cite{rueckert1999}, diffusion penalties in learning-based
registration \cite{balakrishnan2019}, isotropic total variation, which admits sliding without masks
\cite{vishnevskiy2017}, Jacobian-determinant penalties for volume preservation \cite{rohlfing2003}
and against folding \cite{mok2020}, control-point injectivity bounds \cite{choi2000}, diffeomorphic
parameterizations \cite{arsigny2006,ashburner2007,dalca2019}, and hyperelastic regularization \cite{burger2013}. There are two barriers to directly applying this experience in the 2D--3D setting. First, the data term constrains the
motion only through line integrals, so unobserved motion is global rather than local; and the motion basis is itself estimated from those projections.

Here, holding the \emph{VM-DS} optimization framework constant, we explore a range of regularizers and evaluate the performance of each against both training time and accuracy. Our contributions are as follows: (i) We propose the hyperelastic volume term as an efficient replacement for self-consistency. (ii) We compare nine
regularizers, inside one 2D--3D pipeline, with weights and solver chosen on development
cases by rules fixed in advance and tested on eight held-out patients at matched training time.
(iii) We explore the effects of enforcing self-consistency. (iv) We propose a descent-only update rule, which renders the hyperelastic term stable.

\section{Methods}

\subsection{Problem statement}
Given a reference 3D image $\Iref : \Omega \subset \mathbb{R}^3 \to \mathbb{R}$ and target 3D image
$I(\cdot, p)$ on shared grid $\Omega$, we wish to obtain the displacement field $\uu(\x, p)$ such that
\begin{equation}
  \Iref(\x + \uu(\x, p)) \approx I(\x, p)
  \label{eq:problem}
\end{equation}
for all $\x \in \Omega$. Here $\Iref$ is an inspiration-hold CBCT and $I(\cdot, p)$ is
the volumetric image at the time projection $p$ is acquired, which differs from $\Iref$
predominantly by respiratory motion. Only the 2D projection $y_p = P_{\theta_p}[I(\cdot, p)]$ is
measured, where $P_\theta$ is the x-ray forward projector for gantry angle $\theta$. Matching the
projections alone leaves $\uu$ underdetermined, as many fields yield the same $y_p$. The motion model is therefore fit to a regularized objective,
\begin{equation}
  \mathcal{L} = \sum_{p} \rho\!\left(P_{\theta_p}[\Iref(\x + \uu(\x, p))] - y_p\right)
    + \lambda\, \frac{1}{|\Omega|}\int_\Omega r(\uu(\cdot, p))\, d\x ,
  \label{eq:objective}
\end{equation}
where $\rho$ is the pseudo-Huber loss and $r$ a non-negative per-voxel penalty. This paper is
concerned with the choice of $r$. A suitable regularizer should keep the fields physically plausible
(invertible, without folds), should not bias the motion it can express, and should contribute minimally to training time.

\subsection{Baseline method}
\emph{VM-DS} \cite{vmds} fits a low-rank B-spline motion model to a ventilated CBCT arc. Each motion state is described by nine scores $\w(p) = (w_{k,i})$, for axes $k \in \{x,y,z\}$ and three B-spline lattices $i$ such that
\begin{equation}
  u_k(\x, p) = \sum_{i=1}^{3} w_{k,i}(p)\, e_{k,i}(\x),
  \label{eq:model}
\end{equation}
where the spatial components $e_{k,i}$ are shared across projections and parameterized by control
points $C$. Since~\eqref{eq:model} is linear in $\w$ with $C$ fixed and in $C$ with $\w$ fixed,
stage 1 minimises~\eqref{eq:objective} by alternation over the solved projections wherein the w-step consists of one damped Gauss--Newton iteration per projection score, with the penalty entering as residual
rows $\sqrt{\lambda r / |\Omega|}$, and in the c-step the control points take Adam steps on
$\mathcal{L}$ itself, with the basis renormalized to unit RMS displacement. Stage 2 then trains a
convolutional encoder to yield scores from projections. This baseline method uses a self-consistency regularizer,
\begin{equation}
  r = \big(\Iref(\x + \uu + \uu^{-1}\circ(\x + \uu)) - \Iref(\x)\big)^2 ,
  \label{eq:lsc}
\end{equation}
wherein the reference is warped forward, warped back with the
fixed-point inverse \cite{chen2008}.

\subsection{Descent-only c-step}
As published, the c-step of \emph{VM-DS} takes a fixed number of Adam steps and accepts every one. Here, we propose a descent-only variant that re-evaluates $\mathcal{L}$ after each step, undoes any step that raised it by restoring the parameters and optimizer state and halving the learning rate, and on acceptance doubles the rate
back towards its initial value. With this approach, every step of stage 1 descends~\eqref{eq:objective}. The
c-step rule is selected per regularizer (Section~\ref{sec:protocol}).

\subsection{Regularizers compared}
Table~\ref{tab:arms} lists the regularizers, grouped by family:

\begin{itemize}
\item \textbf{Image space.} Self-consistency~\eqref{eq:lsc}, the \emph{VM-DS} baseline \cite{vmds},
warps $\Iref$ forward by $\uu$, warps the result back with the fixed-point inverse \cite{chen2008} and
penalizes the intensity difference from $\Iref$.
\item \textbf{Parameterization.} The stationary velocity field (SVF) used by \emph{VoxelMap} treats
$\uu$ as a velocity and integrates it by scaling and squaring, yielding diffeomorphic warps by
construction \cite{hindley2025voxelmap,dalca2019}.
\item \textbf{Quadratic smoothness.} Bending energy penalizes the second derivatives of the field and
is the classical regularizer for free-form B-spline registration \cite{rueckert1999}. Diffusion
penalizes its first derivatives \cite{balakrishnan2019}. First-order inverse consistency tests the
same round trip as self-consistency, but on the field alone. Since
$\uu^{-1} = -\uu + (\nabla\uu)\,\uu + O(\uu^3)$, the term $\|(\nabla\uu)\,\uu\|^2$ is the error of taking
$-\uu$ as the inverse, and needs no image, no inverse field and no resampling.
\item \textbf{Total variation.} Isotropic TV penalizes gradient magnitude, thereby tolerating discontinuities in the field and has been proposed for sliding motion at the
pleura \cite{vishnevskiy2017}.
\item \textbf{Jacobian.} The negative-Jacobian hinge penalizes folded voxels only \cite{mok2020}.
$|\log J|$ penalizes expansion and compression symmetrically \cite{rohlfing2003}. The control-point
bound projects the control points onto a sufficient condition for injectivity of a B-spline field
\cite{choi2000}. The hyperelastic term penalizes the local change in volume:

\begin{equation}
  r(\uu) = \psi\big(\det J(\x)\big), \qquad \psi(v) = \frac{(v-1)^4}{v^2}, \qquad J = \mathbb{I} + \nabla\uu.
  \label{eq:hyper}
\end{equation}Near $\det J = 1$, $\psi \approx (\det J - 1)^4$, so small
volume changes yield small penalties, while $\psi \to \infty$ as $\det J \to 0$, so compression towards a fold is
heavily penalized.
\end{itemize}

\begin{table}[!htbp]
\centering\small
\caption{Regularization terms. $\eta = 10^{-4}$. Below $J = 0.1$, $\log J$ and $\psi$ are continued by their first-
and second-order Taylor expansions, so folded fields receive a large but finite penalty. $h$ is the
knot spacing.}
\begin{tabularx}{\linewidth}{llX}
\toprule
family & regularizer & per-voxel penalty $r$ \\
\midrule
--- & none & $0$ \\
image space & self-consistency (baseline) & as published, two resamplings, Chen inverse \cite{chen2008} \\
parameterization & SVF & $\uu$ is a velocity and the warp $\exp(\uu)$, 10 squarings \cite{hindley2025voxelmap,dalca2019} \\
quadratic & bending & $\sum_k \big[\sum_a (\partial_{aa}u_k)^2 + 2\sum_{a<b}(\partial_{ab}u_k)^2\big]$ \cite{rueckert1999} \\
 & diffusion & $\sum_{k,a} (\partial_a u_k)^2$ \cite{balakrishnan2019} \\
 & inverse consistency & $\|(\nabla\uu)\,\uu\|^2$, the first-order inverse error \\
total variation & isotropic TV & $\sum_k \sqrt{\|\nabla u_k\|^2 + \eta}$ \cite{vishnevskiy2017} \\
Jacobian & negative-$J$ hinge & $\max(0, -\det J)$ \cite{mok2020} \\
 & $|\log J|$ & $\sqrt{\log^2 \det J + \eta}$ \cite{rohlfing2003} \\
 & control-point bound & projection onto the injectivity bound $0.40\,h$, strict ($\times1$) and relaxed ($\times2$) \cite{choi2000} \\
 & hyperelastic & $\psi(\det J)$, $\psi(v) = (v-1)^4/v^2$ \cite{burger2013} \\
\bottomrule
\end{tabularx}
\label{tab:arms}
\end{table}
\subsection{Data}\label{sec:protocol} We used the public DIR-Lab 4DCT dataset for development and testing \cite{castillo2009}. Each phase volume was resampled to $1 \times 1 \times 1$~mm$^3$, cropped to a single box of $288$~(LR) $\times\ 320$~(SI) $\times\ 248$~(AP)~mm sized to contain every annotated DIR-Lab landmark, and downsampled to $2 \times 2 \times 2$~mm$^3$. The documented end-inhale phase represents the inspiration-hold image and supplies $\bm{I}_{\mathrm{ref}}$. The motion-inclusive projection set $\bm{p}$ is simulated as a circular cone-beam arc of $680$ projections over $360^{\circ}$, with source-to-axis and source-to-detector distances of $1000$ and $1500$~mm and a $272 \times 320$ detector at $1.8$~mm pitch, so that one detector pixel subtends $1.2$~mm at isocentre. A single-breath scan was modelled as a triangular cycle across the arc, with each projection rendered from the phase volume of its respiratory bin by ray-driven forward projection.

Cases 1 and 8 were used for development as they represent the lowest- and highest-motion cases, respectively, while cases 2--7, 9 and 10 were held out. The hyperparameters were selected and evaluated under four rules. First, each $\lambda$ is the value with the lowest mean held-out-view TRE over the two development cases. Second, each c-step rule is selected in the same way. Third, every regularizer that reaches an interior optimum and improves on no regularizer on the development cases is run on the held-out cases, together with \emph{VM-DS}, no regularizer and SVF. Fourth, no hyperparameter is tuned on held-out cases.

\subsection{Experiments} Three experiments were run to understand the effect of each regularizer on training time and accuracy. First, we measured accuracy with a fixed training budget. Stage 1 is capped at 2, 4.25,
8.5 (published) and 17~min, that is, a quarter, a half, one and two times the published budget. Second, we measured training time for a fixed number of iterations. Each regularizer runs exactly 14 outer iterations, matching the published \emph{VM-DS} baseline. A third experiment probed the effect of solving on fewer projections for the best-performing regularizer from the first two experiments. Stage 1 solves on every 30th projection (23 views) instead of every
12th (57), as in the stride-30 configuration of \emph{VM-DS} \cite{vmds}, with stage 1 capped at 1, 2, 4.25
and 8.5~min. Fewer views make each outer iteration cheaper, which shifts the best number of basis
steps per iteration, $K$. We therefore re-selected $K$ for each budget on the development cases and then fixed it
for the held-out cases. Stage 2 amortization is unchanged throughout.

\subsection{Evaluation} Let $\{\x_n\}_{n=1}^{N_L}$ be the landmarks annotated on the reference
and $\{\x_n^b\}$ their annotated positions at phase $b$. A feature at $\x_n$ in the reference appears
at $\x_n + \uu^{-1}(\x_n, p)$, so the error per landmark and view is
\begin{equation}
  \mathrm{TRE}_n(p) = \big\| \x_n + \uu^{-1}(\x_n, p) - \x_n^{b(p)} \big\|_2 ,
  \label{eq:tre}
\end{equation}
where $b(p)$ is the phase of projection $p$ and $\uu$ is evaluated at the encoder-predicted scores.
TRE is averaged over the 75 landmarks (phases T10--T50) and the views of a phase, and then over
phases. Image similarity between the predicted volume, $\Iref$ warped by the predicted field, and
the ground-truth phase volume is assessed by SSIM and PSNR on the 2~mm grid over all nine moving
phases, within a body mask from $\Iref$ \cite{vmds}. Folding is the fraction of
voxels at which the field is not locally invertible,
\begin{equation}
  F = \frac{1}{|\Omega|}\,\big|\{\x \in \Omega : \det J(\x) \le 0\}\big| ,
  \label{eq:fold}
\end{equation}
reported together with percentiles of $\det J$ in soft tissue and parenchyma selected by intensity
alone, for solved fields and for encoder-predicted fields at every held-out view. Regional gain,
the recovered over the true landmark displacement, is reported at the chest wall, the mediastinum
and the diaphragm, with a lung segmentation used only to classify landmarks. Training time is
stage-1 time per outer iteration and the number of outer iterations completed in the budget.
Regularizers are compared with the baseline method by the two-sided Wilcoxon signed-rank test over the eight held-out cases.

\subsection{Implementation details} To isolate the effect of the regularizer, the projection
encoder, the B-spline decoder and the optimizer settings are held fixed \cite{vmds}. In
stage 1, the w-step uses batches of 32 projections and the c-step an Adam learning rate of
$3\times10^{-3}$. In stage 2 the encoder is trained for 100 epochs on 4096 synthesized pairs by Adam
at a learning rate of $10^{-3}$. The encoder applies six layers of $3 \times 3$ convolution, each followed by batch normalization and ReLU, with $2$, $4$, $8$, $16$, $32$ and $32$ channels, followed by a linear layer from the $32 \times 5 \times 5$ features to the nine scores. The decoder has three spatial levels of uniform cubic B-splines, one motion basis component per level, the lattice doubling at each level with $8$, $14$ and $26$ control points per axis at the three levels. Below $\det J = 0.1$, $\log J$ and
$\psi$ are continued by their first- and second-order Taylor expansions, so that folded fields
receive a large but finite penalty. All experiments were run on a workstation with an NVIDIA RTX PRO
4000 Blackwell GPU and an AMD Ryzen Threadripper PRO 9965WX CPU, using PyTorch 2.7
\cite{paszke2019} and the ASTRA toolbox 2.5 \cite{vanaarle2015,vanaarle2016} for forward projection.

\section{Results}
\label{sec:results}

\subsection{Development cases}
\begin{figure}[!htbp]\centering
\IfFileExists{figures/lambda_sweep.pdf}{\includegraphics[width=\linewidth]{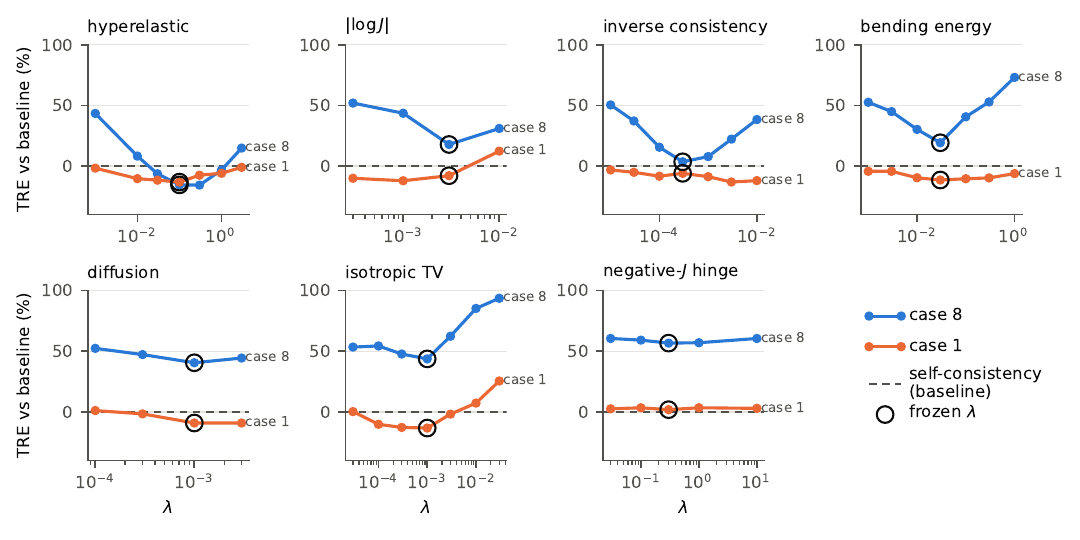}}{\tbd{figure pending}}
\caption{Development sweep. TRE relative to the baseline (self-consistency) on the same case, against
each regularization parameter $\lambda$. Rings indicate the
$\lambda$ frozen for the held-out cases.}
\label{fig:sweep}
\end{figure}

Every regularizer reached an interior optimum
(Figure~\ref{fig:sweep}), and Table~\ref{tab:dev} lists each regularizer at its frozen setting. The hyperelastic term gave the lowest TRE on both
development cases (3.4 and 1.2~mm, against 4.0 and 1.4~mm for the baseline), with a mean of 2.3~mm against 2.7~mm.
$|\log J|$ came next (3.8 and 1.3~mm), also improving on the baseline on both cases, while inverse
consistency was within 0.2~mm of the baseline. The negative-$J$
hinge was no better than no regularizer at any weight over two decades, and the control-point bound
and SVF were worse than no regularizer. Per iteration, the hyperelastic term took 23~s, against 21~s with no regularizer and
35~s with self-consistency, so 21 instead of 14 iterations fit within the 8.5-min budget.
\begin{table}[!htbp]\centering\small\setlength{\tabcolsep}{4pt}
\caption{Development cases. TRE (mm) at each regularizer's frozen setting (selected by development TRE), mean $\pm$ standard deviation over the moving phases.}
\begin{tabular}{lllcc}\toprule
regularizer & $\lambda$ & c-step & case 8 & case 1 \\\midrule
none & --- & desc. & 6.2 $\pm$ 3.0 & 1.4 $\pm$ 0.5  \\
self-consistency (baseline) & --- & Adam & 4.0 $\pm$ 1.3 & 1.4 $\pm$ 0.6  \\
SVF & --- & desc. & 6.9 $\pm$ 3.2 & 1.5 $\pm$ 0.6  \\
bending & $3\!\times\!10^{-2}$ & desc. & 4.8 $\pm$ 2.1 & {\boldmath\bfseries 1.2 $\pm$ 0.4}  \\
diffusion & $10^{-3}$ & Adam & 5.2 $\pm$ 2.3 & 1.4 $\pm$ 0.4  \\
inverse consistency & $3\!\times\!10^{-4}$ & desc. & 4.2 $\pm$ 1.7 & 1.3 $\pm$ 0.4  \\
isotropic TV & $10^{-3}$ & Adam & 5.3 $\pm$ 2.4 & 1.3 $\pm$ 0.5  \\
neg.-$J$ hinge$^\dagger$ & $3\!\times\!10^{-1}$ & desc. & 6.3 $\pm$ 2.8 & 1.4 $\pm$ 0.5  \\
$|\log J|$ & $3\!\times\!10^{-3}$ & Adam & 3.8 $\pm$ 1.4 & 1.3 $\pm$ 0.5  \\
control-point bound$^\dagger$ & $2$ & desc. & 6.9 $\pm$ 3.4 & 1.3 $\pm$ 0.5  \\
{\boldmath\bfseries hyperelastic} & $10^{-1}$ & desc. & {\boldmath\bfseries 3.4 $\pm$ 1.2} & {\boldmath\bfseries 1.2 $\pm$ 0.3}  \\
\bottomrule\end{tabular}
\label{tab:dev}\end{table}

\subsection{Basis-update rule}
With the Adam c-step, the hyperelastic term ($\lambda = 0.1$) descended for 16 outer iterations
on case 8 and then diverged, as the residual rose from 0.19 to 1.29 and the RMS score from 1.0 to
2.9, giving a held-out TRE of 8.3~mm against 3.4~mm with the descent-only c-step.
Self-consistency showed the opposite behavior, with 4.0~mm under the Adam c-step against
5.8~mm under the descent-only one. Without any regularizer, the Adam c-step diverged outright,
whereas the descent-only c-step kept it stable, at 6.2~mm. Across the other regularizers the preference was
mixed (Table~\ref{tab:e2}).$|\log J|$, diffusion and TV were better with the Adam c-step, and
bending energy and inverse consistency marginally better with the descent-only c-step. $|\log J|$ with
the Adam c-step (3.8 and 1.3~mm) also improved on the baseline on both development cases.
\begin{table}[!htbp]\centering\small
\caption{Basis-update rule on the development cases. TRE (mm) at each regularizer's frozen $\lambda$, mean $\pm$ standard deviation over the moving phases. The rule with the lower mean is used on the held-out cases.}
\begin{tabular}{llcccc}\toprule
 & & \multicolumn{2}{c}{Adam c-step} & \multicolumn{2}{c}{descent-only c-step} \\
\cmidrule(lr){3-4}\cmidrule(lr){5-6}
regularizer & $\lambda$ & case 8 & case 1 & case 8 & case 1 \\\midrule
none & --- & diverged & diverged & 6.2 $\pm$ 3.0 & 1.4 $\pm$ 0.5 \\
self-consistency (baseline) & --- & 4.0 $\pm$ 1.3 & 1.4 $\pm$ 0.6 & 5.8 $\pm$ 2.4 & 1.4 $\pm$ 0.5 \\
SVF & --- & 6.7 $\pm$ 3.0 & 1.6 $\pm$ 0.7 & 6.9 $\pm$ 3.2 & 1.5 $\pm$ 0.6 \\
bending & $3\!\times\!10^{-2}$ & 4.9 $\pm$ 2.2 & 1.4 $\pm$ 0.5 & 4.8 $\pm$ 2.1 & {\boldmath\bfseries 1.2 $\pm$ 0.4} \\
diffusion & $10^{-3}$ & 5.2 $\pm$ 2.3 & 1.4 $\pm$ 0.4 & 5.7 $\pm$ 2.6 & 1.3 $\pm$ 0.4 \\
inverse consistency & $3\!\times\!10^{-4}$ & 4.1 $\pm$ 1.6 & 1.5 $\pm$ 0.4 & 4.2 $\pm$ 1.7 & 1.3 $\pm$ 0.4 \\
isotropic TV & $10^{-3}$ & 5.3 $\pm$ 2.4 & {\boldmath\bfseries 1.3 $\pm$ 0.5} & 5.8 $\pm$ 2.6 & {\boldmath\bfseries 1.2 $\pm$ 0.5} \\
$|\log J|$ & $3\!\times\!10^{-3}$ & {\boldmath\bfseries 3.8 $\pm$ 1.4} & {\boldmath\bfseries 1.3 $\pm$ 0.5} & 4.7 $\pm$ 2.0 & 1.3 $\pm$ 0.5 \\
{\boldmath\bfseries hyperelastic} & $10^{-1}$ & 8.3 $\pm$ 4.3 & {\boldmath\bfseries 1.3 $\pm$ 0.4} & {\boldmath\bfseries 3.4 $\pm$ 1.2} & {\boldmath\bfseries 1.2 $\pm$ 0.3} \\
\bottomrule\end{tabular}
\label{tab:e2}\end{table}

\subsection{Held-out patients}
\begin{figure}[!htbp]\centering
\IfFileExists{figures/heldout_diff.pdf}{\includegraphics[width=\linewidth]{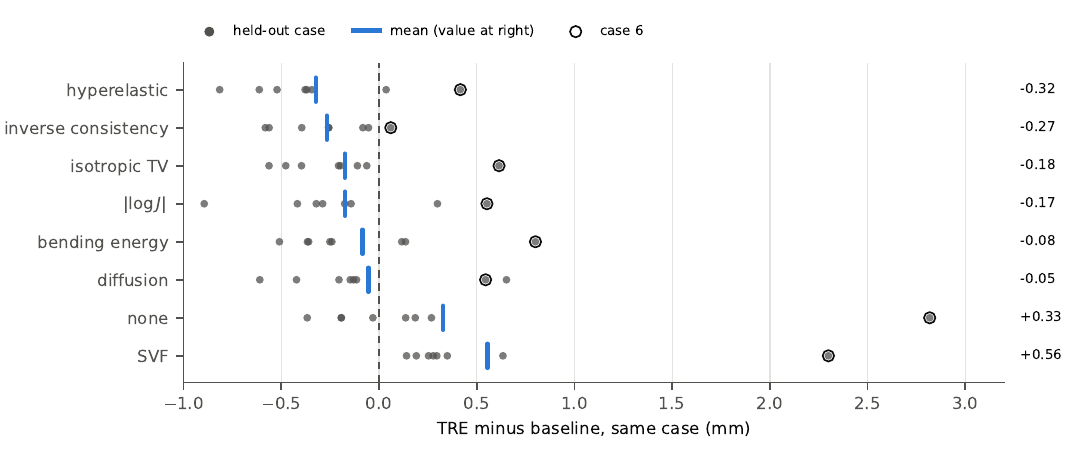}}{\tbd{figure pending}}
\caption{Held-out cases. TRE minus the baseline (self-consistency) on the same case; left of zero is better. Dots indicate cases; bars indicate means (value at right); ringed indicates case~6
(largest motion).}
\label{fig:heldout}
\end{figure}

The baseline reproduced the published held-out TRE of \emph{VM-DS} \cite{vmds}, with a mean of 2.7~mm against
2.8~mm, within 0.1~mm on every case.
Table~\ref{tab:heldout-arc} and Figure~\ref{fig:heldout} compare the regularizers
at their frozen settings. The hyperelastic term had the lowest mean, 2.4~mm ($-0.3$~mm, $-12\%$), and was better than the baseline on six
of eight cases, by up to 0.8~mm. Inverse consistency followed with 2.5~mm, better on
seven of eight cases, and $|\log J|$ and TV gave 2.6~mm each. Bending energy and diffusion were within 0.1~mm of the baseline. Without a
regularizer, TRE was no better than the baseline (3.1~mm), and SVF was worse on every case (3.3~mm, $p = 0.01$). Case~6, the held-out case with the most motion (7.4~mm uncorrected), was the exception. Every geometric regularizer was worse than the baseline.

\begin{table}[!htbp]\centering\small\setlength{\tabcolsep}{3pt}
\caption{Held-out DIR-Lab cases. TRE (mm), at the settings frozen on the development cases. Per case, mean $\pm$ standard deviation over the moving phases; the mean row is the mean $\pm$ standard deviation over the eight cases. The vs rows give the paired mean difference against the baseline (self-consistency), with the two-sided Wilcoxon signed-rank $p$ over the eight cases, uncorrected, in parentheses.}
\resizebox{\linewidth}{!}{%
\begin{tabular}{lcccccccccc}\toprule
case & \shortstack{no\\correction} & none & \shortstack{self-consistency\\(baseline)} & SVF & bending & diffusion & \shortstack{inverse\\consistency} & \shortstack{isotropic\\TV} & $|\log J|$ & {\boldmath\bfseries \shortstack{hyper-\\elastic}} \\\midrule
2 & 3.8 $\pm$ 1.1 & 1.7 $\pm$ 0.1 & 1.9 $\pm$ 0.3 & 2.1 $\pm$ 0.5 & 1.6 $\pm$ 0.2 & 1.7 $\pm$ 0.1 & 1.5 $\pm$ 0.0 & 1.7 $\pm$ 0.2 & 1.5 $\pm$ 0.1 & {\boldmath\bfseries 1.3 $\pm$ 0.0} \\
3 & 5.1 $\pm$ 2.1 & 2.1 $\pm$ 0.4 & 2.5 $\pm$ 0.6 & 2.8 $\pm$ 0.9 & 2.0 $\pm$ 0.5 & 1.9 $\pm$ 0.4 & 1.9 $\pm$ 0.3 & 1.9 $\pm$ 0.4 & {\boldmath\bfseries 1.6 $\pm$ 0.2} & 1.7 $\pm$ 0.2 \\
4 & 6.7 $\pm$ 2.6 & 2.7 $\pm$ 0.8 & 2.9 $\pm$ 0.7 & 3.0 $\pm$ 0.6 & 2.5 $\pm$ 0.6 & 2.4 $\pm$ 0.5 & {\boldmath\bfseries 2.3 $\pm$ 0.5} & 2.4 $\pm$ 0.6 & 3.2 $\pm$ 1.0 & 2.5 $\pm$ 0.7 \\
5 & 5.2 $\pm$ 2.0 & 3.1 $\pm$ 0.8 & 2.9 $\pm$ 0.7 & 3.1 $\pm$ 0.7 & {\boldmath\bfseries 2.5 $\pm$ 0.4} & 2.8 $\pm$ 0.5 & 2.6 $\pm$ 0.5 & {\boldmath\bfseries 2.5 $\pm$ 0.4} & {\boldmath\bfseries 2.5 $\pm$ 0.4} & {\boldmath\bfseries 2.5 $\pm$ 0.4} \\
6 & 7.4 $\pm$ 3.5 & 6.2 $\pm$ 2.8 & {\boldmath\bfseries 3.4 $\pm$ 0.9} & 5.7 $\pm$ 2.5 & 4.2 $\pm$ 1.4 & 3.9 $\pm$ 1.5 & 3.5 $\pm$ 1.0 & 4.0 $\pm$ 1.5 & 3.9 $\pm$ 1.2 & 3.8 $\pm$ 1.2 \\
7 & 6.7 $\pm$ 3.1 & 3.4 $\pm$ 0.9 & 3.2 $\pm$ 0.9 & 3.8 $\pm$ 1.1 & 3.3 $\pm$ 1.0 & 3.9 $\pm$ 1.2 & 3.0 $\pm$ 0.7 & 3.1 $\pm$ 0.9 & 3.0 $\pm$ 0.4 & {\boldmath\bfseries 2.6 $\pm$ 0.6} \\
9 & 5.0 $\pm$ 2.2 & 2.5 $\pm$ 0.4 & 2.4 $\pm$ 0.6 & 2.7 $\pm$ 0.7 & 2.5 $\pm$ 0.6 & {\boldmath\bfseries 2.2 $\pm$ 0.4} & 2.3 $\pm$ 0.4 & {\boldmath\bfseries 2.2 $\pm$ 0.5} & {\boldmath\bfseries 2.2 $\pm$ 0.4} & 2.4 $\pm$ 0.3 \\
10 & 5.4 $\pm$ 2.6 & 2.9 $\pm$ 0.8 & 2.9 $\pm$ 0.8 & 3.2 $\pm$ 0.9 & 2.7 $\pm$ 0.8 & 2.7 $\pm$ 0.7 & 2.9 $\pm$ 0.7 & 2.7 $\pm$ 0.7 & {\boldmath\bfseries 2.6 $\pm$ 0.6} & {\boldmath\bfseries 2.6 $\pm$ 0.5} \\\midrule
mean & 5.7 $\pm$ 1.1 & 3.1 $\pm$ 1.3 & 2.7 $\pm$ 0.5 & 3.3 $\pm$ 1.0 & 2.7 $\pm$ 0.7 & 2.7 $\pm$ 0.8 & 2.5 $\pm$ 0.6 & 2.6 $\pm$ 0.7 & 2.6 $\pm$ 0.8 & {\boldmath\bfseries 2.4 $\pm$ 0.7} \\
\quad vs baseline &  & +0.3 (0.95) &  & +0.6 (0.01) & $-$0.1 (0.38) & $-$0.1 (0.64) & $-$0.3 (0.02) & $-$0.2 (0.20) & $-$0.2 (0.38) & $-$0.3 (0.11) \\
\bottomrule\end{tabular}}
\label{tab:heldout-arc}\end{table}

\begin{figure}[p]\centering
\IfFileExists{figures/images_c2.pdf}{\includegraphics[width=0.94\linewidth]{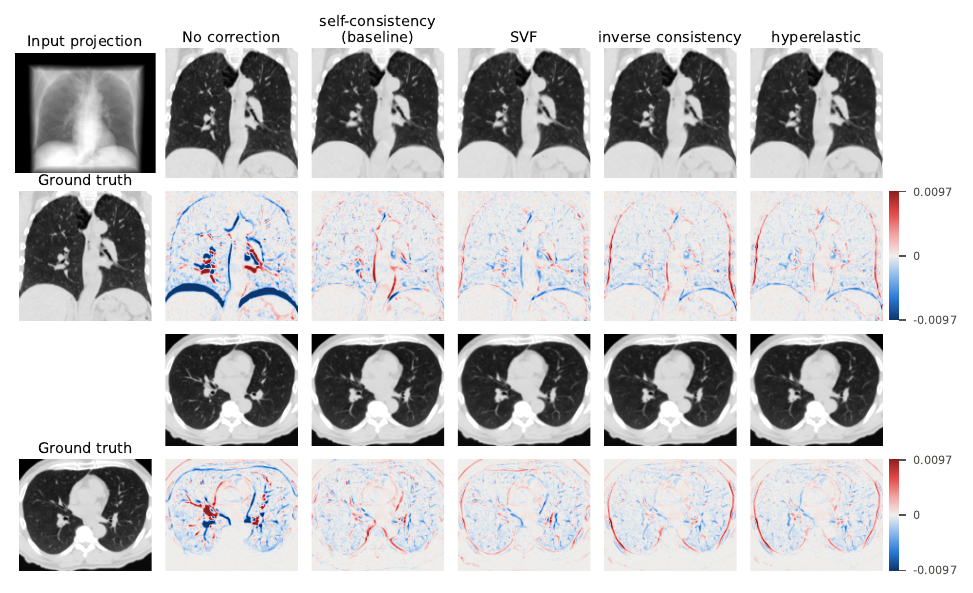}}{\tbd{figure pending}}
\caption{The held-out case on which the hyperelastic term was most accurate, case~2, at end-exhale.
In each pair, the upper row shows the inspiration-hold reference (no correction) and each predicted
volume, and the lower row the ground truth, then each volume minus the ground truth.}
\label{fig:images}

\bigskip
\IfFileExists{figures/images_c6.pdf}{\includegraphics[width=0.94\linewidth]{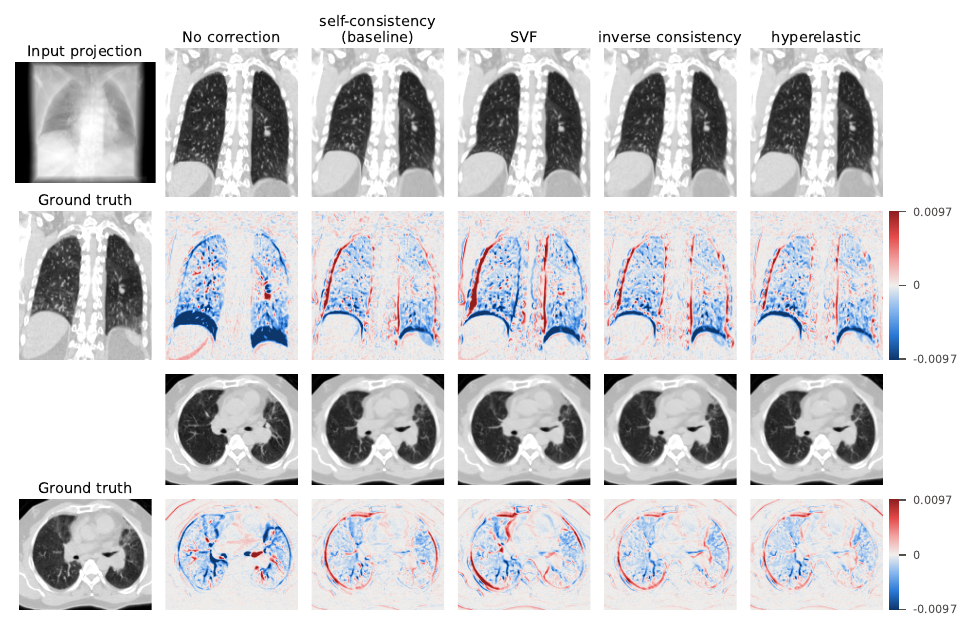}}{\tbd{figure pending}}
\caption{As Figure~\ref{fig:images}, for the held-out case on which the hyperelastic term was least
accurate, case~6, which also has the most motion.}
\label{fig:images-worst}
\end{figure}

Table~\ref{tab:heldout-summary} reports SSIM and PSNR of the predicted volumes, with per-case values
in Tables~\ref{tab:supp-ssim-arc} and~\ref{tab:supp-psnr-arc}. The baseline reproduced the published values of \emph{VM-DS} (SSIM 0.95, PSNR 37.7~dB \cite{vmds}). The geometric penalties were within 0.01 of the baseline in SSIM and 0.3~dB in PSNR, with
isotropic TV and inverse consistency slightly higher in SSIM (0.96 and 0.96 against
0.95, $p = 0.01$ and $0.04$). The regularizers that were less accurate in TRE were also lower in image
similarity, SVF by 1.5~dB and no regularizer by 0.6~dB ($p = 0.01$ and $0.02$). Figures~\ref{fig:images}
and~\ref{fig:images-worst} show the predicted volumes for the held-out cases on which the
hyperelastic term was most and least accurate.
\begin{table}[!htbp]\centering\small\setlength{\tabcolsep}{2pt}
\caption{Held-out cases, mean $\pm$ standard deviation over the eight cases. TRE, and SSIM and PSNR of the predicted volume against the ground-truth phase volume. In brackets, the paired mean difference against the baseline (self-consistency), with the two-sided Wilcoxon signed-rank $p$ over the eight cases, uncorrected, in parentheses.}
\begin{tabular}{lccc}\toprule
regularizer & TRE (mm) & SSIM & PSNR (dB) \\\midrule
no correction & 5.7 $\pm$ 1.1 & 0.92 $\pm$ 0.02 & 33.5 $\pm$ 3.7  \\\midrule
none  & 3.1 $\pm$ 1.3 \,[+0.3 (0.95)] & 0.95 $\pm$ 0.02 \,[$-$0.01 (0.02)] & 37.1 $\pm$ 4.3 \,[$-$0.5 (0.02)] \\
self-consistency (baseline) & 2.7 $\pm$ 0.5 & 0.95 $\pm$ 0.02 & 37.7 $\pm$ 4.0  \\
SVF  & 3.3 $\pm$ 1.0 \,[+0.6 (0.01)] & 0.94 $\pm$ 0.02 \,[$-$0.01 (0.01)] & 36.2 $\pm$ 4.2 \,[$-$1.5 (0.01)] \\
bending  & 2.7 $\pm$ 0.7 \,[$-$0.1 (0.38)] & 0.95 $\pm$ 0.02 \,[+0.00 (0.64)] & 37.5 $\pm$ 4.3 \,[$-$0.2 (0.38)] \\
diffusion  & 2.7 $\pm$ 0.8 \,[$-$0.1 (0.64)] & 0.95 $\pm$ 0.02 \,[+0.00 (0.95)] & 37.4 $\pm$ 4.4 \,[$-$0.2 (0.25)] \\
inverse consistency  & 2.5 $\pm$ 0.6 \,[$-$0.3 (0.02)] & {\boldmath\bfseries 0.96 $\pm$ 0.02} \,[+0.00 (0.04)] & 37.8 $\pm$ 4.2 \,[+0.2 (0.31)] \\
{\boldmath\bfseries isotropic TV}  & 2.6 $\pm$ 0.7 \,[$-$0.2 (0.20)] & {\boldmath\bfseries 0.96 $\pm$ 0.02} \,[+0.00 (0.01)] & {\boldmath\bfseries 37.9 $\pm$ 4.3} \,[+0.3 (0.38)] \\
{\boldmath\bfseries $|\log J|$}  & 2.6 $\pm$ 0.8 \,[$-$0.2 (0.38)] & {\boldmath\bfseries 0.96 $\pm$ 0.02} \,[+0.00 (0.08)] & {\boldmath\bfseries 37.9 $\pm$ 4.2} \,[+0.2 (0.38)] \\
{\boldmath\bfseries hyperelastic}  & {\boldmath\bfseries 2.4 $\pm$ 0.7} \,[$-$0.3 (0.11)] & {\boldmath\bfseries 0.96 $\pm$ 0.02} \,[+0.00 (0.08)] & 37.8 $\pm$ 4.2 \,[+0.1 (0.64)] \\
\bottomrule\end{tabular}
\label{tab:heldout-summary}\end{table}
\begin{tablepair}
\begin{table}[!htbp]\centering\small\setlength{\tabcolsep}{3pt}
\caption{Held-out DIR-Lab cases. SSIM of the predicted volume against the ground-truth phase volume. Per case, mean $\pm$ standard deviation over the moving phases; the mean row is the mean $\pm$ standard deviation over the eight cases. The vs rows give the paired mean difference against the baseline (self-consistency), with the two-sided Wilcoxon signed-rank $p$ over the eight cases, uncorrected, in parentheses.}
\resizebox{\linewidth}{!}{%
\begin{tabular}{lcccccccccc}\toprule
case & \shortstack{no\\correction} & none & \shortstack{self-consistency\\(baseline)} & SVF & bending & diffusion & {\boldmath\bfseries \shortstack{inverse\\consistency}} & \shortstack{isotropic\\TV} & $|\log J|$ & \shortstack{hyper-\\elastic} \\\midrule
2 & 0.89 $\pm$ 0.04 & 0.95 $\pm$ 0.01 & 0.95 $\pm$ 0.01 & 0.94 $\pm$ 0.02 & 0.95 $\pm$ 0.01 & 0.95 $\pm$ 0.01 & 0.95 $\pm$ 0.00 & 0.95 $\pm$ 0.01 & 0.95 $\pm$ 0.01 & {\boldmath\bfseries 0.96 $\pm$ 0.01} \\
3 & 0.97 $\pm$ 0.02 & {\boldmath\bfseries 0.99 $\pm$ 0.00} & {\boldmath\bfseries 0.99 $\pm$ 0.00} & {\boldmath\bfseries 0.99 $\pm$ 0.00} & {\boldmath\bfseries 0.99 $\pm$ 0.00} & {\boldmath\bfseries 0.99 $\pm$ 0.00} & {\boldmath\bfseries 0.99 $\pm$ 0.00} & {\boldmath\bfseries 0.99 $\pm$ 0.00} & {\boldmath\bfseries 0.99 $\pm$ 0.00} & {\boldmath\bfseries 0.99 $\pm$ 0.00} \\
4 & 0.92 $\pm$ 0.03 & 0.95 $\pm$ 0.01 & {\boldmath\bfseries 0.96 $\pm$ 0.01} & 0.94 $\pm$ 0.01 & {\boldmath\bfseries 0.96 $\pm$ 0.01} & 0.95 $\pm$ 0.01 & {\boldmath\bfseries 0.96 $\pm$ 0.01} & {\boldmath\bfseries 0.96 $\pm$ 0.01} & 0.95 $\pm$ 0.01 & {\boldmath\bfseries 0.96 $\pm$ 0.01} \\
5 & 0.93 $\pm$ 0.03 & 0.96 $\pm$ 0.01 & 0.96 $\pm$ 0.01 & 0.95 $\pm$ 0.01 & {\boldmath\bfseries 0.97 $\pm$ 0.00} & 0.96 $\pm$ 0.01 & {\boldmath\bfseries 0.97 $\pm$ 0.00} & {\boldmath\bfseries 0.97 $\pm$ 0.01} & {\boldmath\bfseries 0.97 $\pm$ 0.01} & 0.96 $\pm$ 0.01 \\
6 & 0.90 $\pm$ 0.04 & 0.92 $\pm$ 0.03 & 0.93 $\pm$ 0.02 & 0.91 $\pm$ 0.03 & 0.93 $\pm$ 0.02 & 0.93 $\pm$ 0.03 & {\boldmath\bfseries 0.94 $\pm$ 0.02} & 0.93 $\pm$ 0.02 & {\boldmath\bfseries 0.94 $\pm$ 0.02} & 0.93 $\pm$ 0.02 \\
7 & 0.89 $\pm$ 0.04 & 0.92 $\pm$ 0.03 & 0.93 $\pm$ 0.02 & 0.91 $\pm$ 0.03 & 0.93 $\pm$ 0.02 & 0.93 $\pm$ 0.02 & {\boldmath\bfseries 0.94 $\pm$ 0.02} & {\boldmath\bfseries 0.94 $\pm$ 0.01} & {\boldmath\bfseries 0.94 $\pm$ 0.01} & {\boldmath\bfseries 0.94 $\pm$ 0.02} \\
9 & 0.92 $\pm$ 0.03 & {\boldmath\bfseries 0.96 $\pm$ 0.01} & {\boldmath\bfseries 0.96 $\pm$ 0.01} & {\boldmath\bfseries 0.96 $\pm$ 0.01} & {\boldmath\bfseries 0.96 $\pm$ 0.01} & {\boldmath\bfseries 0.96 $\pm$ 0.01} & {\boldmath\bfseries 0.96 $\pm$ 0.01} & {\boldmath\bfseries 0.96 $\pm$ 0.01} & {\boldmath\bfseries 0.96 $\pm$ 0.01} & {\boldmath\bfseries 0.96 $\pm$ 0.01} \\
10 & 0.91 $\pm$ 0.04 & 0.94 $\pm$ 0.02 & 0.94 $\pm$ 0.02 & 0.93 $\pm$ 0.02 & 0.94 $\pm$ 0.01 & 0.94 $\pm$ 0.01 & {\boldmath\bfseries 0.95 $\pm$ 0.01} & {\boldmath\bfseries 0.95 $\pm$ 0.01} & {\boldmath\bfseries 0.95 $\pm$ 0.01} & {\boldmath\bfseries 0.95 $\pm$ 0.02} \\\midrule
mean & 0.92 $\pm$ 0.02 & 0.95 $\pm$ 0.02 & 0.95 $\pm$ 0.02 & 0.94 $\pm$ 0.02 & 0.95 $\pm$ 0.02 & 0.95 $\pm$ 0.02 & {\boldmath\bfseries 0.96 $\pm$ 0.02} & {\boldmath\bfseries 0.96 $\pm$ 0.02} & {\boldmath\bfseries 0.96 $\pm$ 0.02} & {\boldmath\bfseries 0.96 $\pm$ 0.02} \\
\quad vs baseline &  & $-$0.01 (0.02) &  & $-$0.01 (0.01) & +0.00 (0.64) & +0.00 (0.95) & +0.00 (0.04) & +0.00 (0.01) & +0.00 (0.08) & +0.00 (0.08) \\
\bottomrule\end{tabular}}
\label{tab:supp-ssim-arc}\end{table}
\bigskip
\end{tablepair}
\begin{tablepair}
\begin{table}[!htbp]\centering\small\setlength{\tabcolsep}{3pt}
\caption{Held-out DIR-Lab cases. PSNR (dB) of the predicted volume against the ground-truth phase volume. Per case, mean $\pm$ standard deviation over the moving phases; the mean row is the mean $\pm$ standard deviation over the eight cases. The vs rows give the paired mean difference against the baseline (self-consistency), with the two-sided Wilcoxon signed-rank $p$ over the eight cases, uncorrected, in parentheses.}
\resizebox{\linewidth}{!}{%
\begin{tabular}{lcccccccccc}\toprule
case & \shortstack{no\\correction} & none & \shortstack{self-consistency\\(baseline)} & SVF & bending & diffusion & \shortstack{inverse\\consistency} & {\boldmath\bfseries \shortstack{isotropic\\TV}} & $|\log J|$ & \shortstack{hyper-\\elastic} \\\midrule
2 & 28.8 $\pm$ 3.6 & 33.8 $\pm$ 0.8 & {\boldmath\bfseries 34.4 $\pm$ 1.3} & 32.5 $\pm$ 1.4 & 33.6 $\pm$ 1.4 & 33.5 $\pm$ 1.0 & 34.1 $\pm$ 0.7 & 34.2 $\pm$ 1.4 & 34.2 $\pm$ 1.4 & 34.3 $\pm$ 1.0 \\
3 & 42.5 $\pm$ 4.1 & 47.9 $\pm$ 1.3 & 47.8 $\pm$ 1.9 & 46.6 $\pm$ 1.7 & 48.2 $\pm$ 1.8 & 48.4 $\pm$ 1.6 & 48.2 $\pm$ 1.7 & {\boldmath\bfseries 48.7 $\pm$ 1.9} & 48.5 $\pm$ 2.3 & 48.3 $\pm$ 1.5 \\
4 & 33.6 $\pm$ 2.8 & 37.7 $\pm$ 2.2 & 37.9 $\pm$ 1.7 & 36.3 $\pm$ 1.6 & 37.9 $\pm$ 1.9 & 37.5 $\pm$ 1.8 & {\boldmath\bfseries 38.5 $\pm$ 1.9} & 38.3 $\pm$ 2.0 & 37.4 $\pm$ 2.0 & 38.3 $\pm$ 2.1 \\
5 & 33.0 $\pm$ 2.9 & 37.2 $\pm$ 1.7 & 37.5 $\pm$ 1.4 & 36.1 $\pm$ 1.2 & 38.0 $\pm$ 0.7 & 37.6 $\pm$ 1.3 & 38.1 $\pm$ 0.9 & {\boldmath\bfseries 38.5 $\pm$ 1.3} & {\boldmath\bfseries 38.5 $\pm$ 1.2} & 37.9 $\pm$ 1.1 \\
6 & 32.7 $\pm$ 3.5 & 34.5 $\pm$ 2.9 & 35.4 $\pm$ 2.3 & 34.0 $\pm$ 2.6 & 35.1 $\pm$ 2.4 & 35.1 $\pm$ 2.8 & {\boldmath\bfseries 35.8 $\pm$ 2.3} & 35.4 $\pm$ 2.5 & 35.7 $\pm$ 2.3 & 35.4 $\pm$ 2.2 \\
7 & 31.6 $\pm$ 3.0 & 34.1 $\pm$ 2.1 & 34.9 $\pm$ 1.9 & 33.3 $\pm$ 1.9 & 34.9 $\pm$ 1.7 & 34.9 $\pm$ 1.3 & 35.1 $\pm$ 1.6 & 35.1 $\pm$ 1.6 & 35.0 $\pm$ 1.3 & {\boldmath\bfseries 35.3 $\pm$ 1.8} \\
9 & 33.2 $\pm$ 2.9 & 36.7 $\pm$ 1.1 & {\boldmath\bfseries 37.5 $\pm$ 1.1} & 36.6 $\pm$ 1.3 & 36.9 $\pm$ 1.1 & 37.1 $\pm$ 1.1 & 36.9 $\pm$ 0.9 & {\boldmath\bfseries 37.5 $\pm$ 1.2} & 37.4 $\pm$ 1.3 & 37.0 $\pm$ 0.8 \\
10 & 32.3 $\pm$ 3.1 & 35.1 $\pm$ 2.2 & 35.9 $\pm$ 1.6 & 34.0 $\pm$ 1.6 & 35.5 $\pm$ 1.5 & 35.5 $\pm$ 1.6 & 35.9 $\pm$ 1.2 & 35.8 $\pm$ 1.5 & {\boldmath\bfseries 36.3 $\pm$ 1.4} & 35.7 $\pm$ 1.7 \\\midrule
mean & 33.5 $\pm$ 3.7 & 37.1 $\pm$ 4.3 & 37.7 $\pm$ 4.0 & 36.2 $\pm$ 4.2 & 37.5 $\pm$ 4.3 & 37.4 $\pm$ 4.4 & 37.8 $\pm$ 4.2 & {\boldmath\bfseries 37.9 $\pm$ 4.3} & {\boldmath\bfseries 37.9 $\pm$ 4.2} & 37.8 $\pm$ 4.2 \\
\quad vs baseline &  & $-$0.5 (0.02) &  & $-$1.5 (0.01) & $-$0.2 (0.38) & $-$0.2 (0.25) & +0.2 (0.31) & +0.3 (0.38) & +0.2 (0.38) & +0.1 (0.64) \\
\bottomrule\end{tabular}}
\label{tab:supp-psnr-arc}\end{table}
\bigskip
\end{tablepair}

\subsection{Training time}
\label{sec:time}
\begin{figure}[!htbp]\centering
\IfFileExists{figures/time_bar.pdf}{\includegraphics[width=\linewidth]{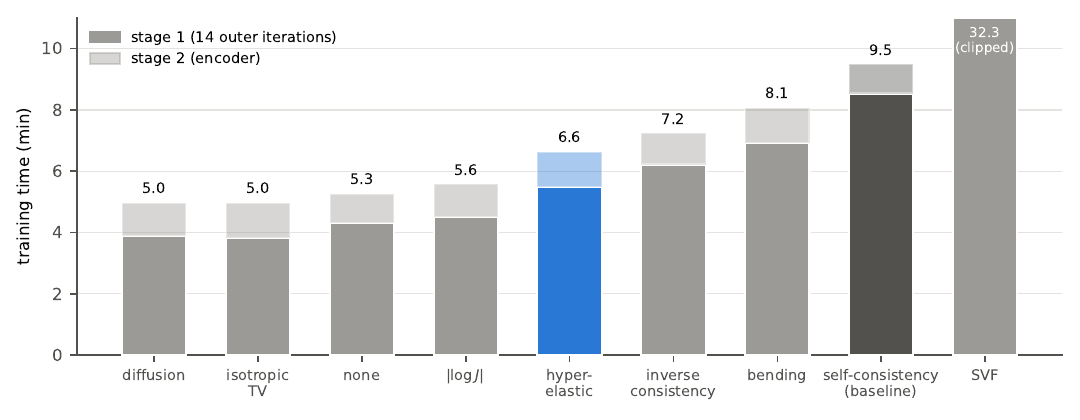}}{\tbd{figure pending}}
\caption{Training time per regularizer, mean over the eight held-out cases. Stage 1 (dark) for 14
outer iterations, the count \emph{VM-DS} completes in its 8.5-min budget, from the time per outer
iteration in the 8.5-min runs; stage 2 (light) stacked on top.}
\label{fig:time}
\end{figure}

Table~\ref{tab:outers} reports the fixed-iteration experiment. Run for exactly 14 outer
iterations, the baseline took 8.4~min of stage-1 training (36~s per iteration) and the hyperelastic term
5.6~min (24~s), 33\% less, while inverse consistency took 6.2~min and no regularizer 4.7~min.
Training time varied by at most 0.4~min between cases for any regularizer. At the same number of iterations,
mean held-out TRE of the hyperelastic term was 2.5 and 2.4~mm against 2.8 and 2.8~mm for the baseline
($p = 0.04$ and $0.05$). Figure~\ref{fig:time} extends the comparison to every regularizer
run on the held-out cases.
\begin{table}[!htbp]\centering\small\setlength{\tabcolsep}{3pt}
\caption{Fixed iterations. Every regularizer runs exactly 14 outer iterations without a cap. Training time: stage 1 over the ten cases. Held-out TRE, mean $\pm$ standard deviation over the eight cases; development cases, mean $\pm$ standard deviation over the moving phases. The vs rows give the paired mean difference against the baseline (self-consistency), with the two-sided Wilcoxon signed-rank $p$ over the eight cases, uncorrected, in parentheses.}
\begin{tabular}{lcccccc}\toprule
 & \multicolumn{3}{c}{stage-1 training time} & \multicolumn{3}{c}{TRE (mm)} \\
\cmidrule(lr){2-4}\cmidrule(lr){5-7}
regularizer & mean (min) & range & s/outer & held-out & case 8 & case 1 \\\midrule
none & 4.7 & 4.5--4.8 & 20 & 3.2 $\pm$ 1.2 & 6.4 $\pm$ 3.1 & 1.4 $\pm$ 0.5 \\
\quad vs baseline &  &  &  & +0.4 (0.55) &  &  \\
self-consistency (baseline) & 8.4 & 8.2--8.6 & 36 & 2.8 $\pm$ 0.5 & 3.9 $\pm$ 1.2 & 1.4 $\pm$ 0.6 \\
inverse consistency & 6.2 & 6.1--6.3 & 27 & 2.6 $\pm$ 0.7 & 4.3 $\pm$ 1.6 & 1.3 $\pm$ 0.4 \\
\quad vs baseline &  &  &  & $-$0.3 (0.08) &  &  \\
{\boldmath\bfseries hyperelastic} & 5.6 & 5.5--5.8 & 24 & {\boldmath\bfseries 2.5 $\pm$ 0.7} & {\boldmath\bfseries 3.7 $\pm$ 1.5} & {\boldmath\bfseries 1.2 $\pm$ 0.4} \\
\quad vs baseline &  &  &  & $-$0.4 (0.04) &  &  \\
\bottomrule\end{tabular}
\label{tab:outers}\end{table}

Table~\ref{tab:budget} reports TRE against the stage-1 budget. The
caps were met to within 0.25~min, with the baseline using 1.8, 4.1 and 16.8~min at caps of 2, 4.25 and
17~min. At the same budget, the hyperelastic term had a lower held-out mean than the baseline at every
budget, by 0.3--0.5~mm, on six to eight of eight cases
($p = 0.01$ to $0.11$, uncorrected). With a quarter of the published budget (2~min; 3.2~min of
training in total, against 9.7~min), it matched the baseline at 8.5~min (2.7 against 2.7~mm,
$p = 0.64$). With half the budget (4.25~min) it was
lower than the baseline at either 8.5 or 17~min (2.6~mm, against at best 2.7~mm), though not significantly. Given more time, the baseline did not improve on the held-out cases
beyond 8.5~min (2.8~mm at 17~min), although case~8 did (3.3~mm, against
4.0~mm). The hyperelastic term kept improving, to 2.3~mm at 17~min, as did inverse
consistency, whereas without a regularizer held-out TRE stayed near 3.1~mm at every
budget. Case~6 remained the exception at 4.25~min and above ($+0.2$ to $+0.4$~mm against the baseline),
but at 2~min the hyperelastic term was better there too.

\begin{table}[!htbp]\centering\small\setlength{\tabcolsep}{4pt}
\caption{Fixed stage-1 budget. TRE (mm). Development cases, mean $\pm$ standard deviation over the moving phases; held-out, mean $\pm$ standard deviation over the eight cases. Outers: outer iterations completed, mean over the ten cases. In brackets, the paired mean difference against the baseline (self-consistency) at the same budget, with the two-sided Wilcoxon signed-rank $p$ over the eight cases, uncorrected, in parentheses.}
\begin{tabular}{llcccc}\toprule
regularizer & budget (min) & case 8 & case 1 & held-out mean & outers \\\midrule
none & 2 & 6.9 $\pm$ 3.4 & 1.4 $\pm$ 0.5 & 3.2 $\pm$ 1.0 \,[+0.1 (0.84)] & 5.5 \\
 & 4.25 & 6.4 $\pm$ 3.2 & 1.4 $\pm$ 0.5 & 3.2 $\pm$ 1.2 \,[+0.3 (0.15)] & 12.2 \\
 & 8.5 & 6.2 $\pm$ 3.0 & 1.4 $\pm$ 0.5 & 3.1 $\pm$ 1.3 \,[+0.3 (0.95)] & 26.6 \\
 & 17 & 6.1 $\pm$ 3.0 & 1.4 $\pm$ 0.5 & 3.1 $\pm$ 1.4 \,[+0.3 (1.00)] & 49.8 \\
self-consistency (baseline) & 2 & 5.6 $\pm$ 2.6 & 1.4 $\pm$ 0.6 & 3.1 $\pm$ 0.8 & 3.0 \\
 & 4.25 & 4.8 $\pm$ 2.0 & 1.4 $\pm$ 0.5 & 2.9 $\pm$ 0.6 & 6.8 \\
 & 8.5 & 4.0 $\pm$ 1.3 & 1.4 $\pm$ 0.6 & 2.7 $\pm$ 0.5 & 13.5 \\
 & 17 & 3.3 $\pm$ 1.0 & 1.5 $\pm$ 0.7 & 2.8 $\pm$ 0.5 & 27.9 \\\midrule
inverse consistency & 2 & 5.2 $\pm$ 2.3 & 1.3 $\pm$ 0.5 & 2.9 $\pm$ 0.8 \,[$-$0.2 (0.04)] & 4.0 \\
 & 4.25 & 4.6 $\pm$ 1.9 & 1.3 $\pm$ 0.5 & 2.7 $\pm$ 0.7 \,[$-$0.2 (0.04)] & 9.0 \\
 & 8.5 & 4.2 $\pm$ 1.7 & 1.3 $\pm$ 0.4 & 2.5 $\pm$ 0.6 \,[$-$0.3 (0.02)] & 18.7 \\
 & 17 & 4.0 $\pm$ 1.5 & 1.3 $\pm$ 0.3 & 2.4 $\pm$ 0.6 \,[$-$0.4 (0.02)] & 37.3 \\
{\boldmath\bfseries hyperelastic} & 2 & {\boldmath\bfseries 4.7 $\pm$ 2.0} & {\boldmath\bfseries 1.2 $\pm$ 0.4} & {\boldmath\bfseries 2.7 $\pm$ 0.7} \,[$-$0.4 (0.01)] & 4.7 \\
 & 4.25 & {\boldmath\bfseries 3.9 $\pm$ 1.6} & {\boldmath\bfseries 1.2 $\pm$ 0.4} & {\boldmath\bfseries 2.6 $\pm$ 0.8} \,[$-$0.3 (0.04)] & 10.0 \\
 & 8.5 & {\boldmath\bfseries 3.4 $\pm$ 1.2} & {\boldmath\bfseries 1.2 $\pm$ 0.3} & {\boldmath\bfseries 2.4 $\pm$ 0.7} \,[$-$0.3 (0.11)] & 21.2 \\
 & 17 & {\boldmath\bfseries 3.2 $\pm$ 1.0} & {\boldmath\bfseries 1.2 $\pm$ 0.3} & {\boldmath\bfseries 2.3 $\pm$ 0.7} \,[$-$0.5 (0.04)] & 42.7 \\
\bottomrule\end{tabular}
\label{tab:budget}\end{table}
\subsection{The number of solved projections}
\label{sec:stride30}
Table~\ref{tab:stride30} reports the hyperelastic term solving on 23 instead of 57 projections.
On the development cases, $K = 23$ was best at 2 to 8.5~min and $K = 14$ at 1~min
(Table~\ref{tab:supp-s30k}). Larger $K$ was worse at every budget, increasingly so as the budget
shrank. On case 8 at 2~min, for example, TRE was 4.8~mm at $K = 56$ (the stride-30 setting of
\emph{VM-DS}) against 3.9~mm at $K = 23$. With stage 1 capped at 1~min (0.93~min used, 9 outer
iterations), held-out TRE was 2.7~mm, against 2.7~mm
for the baseline at its published configuration, which corresponds to about 2~min of training in total
instead of nearly 10. Between 1 and 8.5~min the held-out mean moved by at most 0.2~mm (2.5--2.7~mm). At 2 and 4.25~min it was as accurate as stride 12
at the same budget.
No stride-30 budget differed significantly from the baseline ($p > 0.1$), with four to seven of the eight
cases lower. Case~6 was worse at stride 30 than at stride 12, at 4.2--4.9~mm against
3.8--4.2~mm and 3.4~mm for the baseline, whereas on the other seven cases the stride-30 mean was
2.2--2.5~mm against 2.7~mm for the baseline. Image similarity was within 0.01 SSIM and 0.6~dB PSNR of
the baseline at every budget.
\begin{table}[!htbp]\centering\small\setlength{\tabcolsep}{4pt}
\caption{Accuracy with fewer solved projections. Held-out TRE (mm) of the hyperelastic term at stride 12 (57 views) and 30 (23 views), mean $\pm$ standard deviation over the eight cases; case 6, mean $\pm$ standard deviation over the moving phases. At stride 30, $K$ was selected per budget on the development cases (Table~\ref{tab:supp-s30k}). The vs rows give the paired mean difference against the baseline (last row), with the two-sided Wilcoxon signed-rank $p$ over the eight cases, uncorrected, and the number of cases lower than the baseline, in parentheses.}
\begin{tabular}{llcccc}\toprule
 & & & & \multicolumn{2}{c}{TRE (mm)} \\
\cmidrule(lr){5-6}
stride & budget (min) & $K$ & outers & held-out & case 6 \\\midrule
12 & 8.5 & 23 & 21.2 & {\boldmath\bfseries 2.4 $\pm$ 0.7} & 3.8 $\pm$ 1.2 \\
 & \quad vs baseline &  &  & $-$0.3 (0.11, 6/8) &  \\
 & 4.25 & 23 & 10.0 & 2.6 $\pm$ 0.8 & 4.1 $\pm$ 1.3 \\
 & \quad vs baseline &  &  & $-$0.2 (0.20, 7/8) &  \\
 & 2 & 23 & 4.6 & 2.7 $\pm$ 0.7 & 4.2 $\pm$ 1.2 \\
 & \quad vs baseline &  &  & $-$0.0 (0.64, 4/8) &  \\\midrule
30 & 8.5 & 23 & 52.0 & 2.5 $\pm$ 0.9 & 4.6 $\pm$ 1.7 \\
 & \quad vs baseline &  &  & $-$0.2 (0.20, 7/8) &  \\
 & 4.25 & 23 & 26.0 & 2.6 $\pm$ 0.9 & 4.8 $\pm$ 1.7 \\
 & \quad vs baseline &  &  & $-$0.2 (0.20, 7/8) &  \\
 & 2 & 23 & 12.0 & 2.7 $\pm$ 1.0 & 4.9 $\pm$ 1.8 \\
 & \quad vs baseline &  &  & $-$0.0 (0.20, 7/8) &  \\
 & 1 & 14 & 9.0 & 2.7 $\pm$ 0.7 & 4.2 $\pm$ 1.3 \\
 & \quad vs baseline &  &  & $-$0.1 (0.38, 5/8) &  \\\midrule
\multicolumn{4}{l}{self-consistency (baseline; stride 12, 8.5~min)} & 2.7 $\pm$ 0.5 & {\boldmath\bfseries 3.4 $\pm$ 0.9} \\
\bottomrule\end{tabular}
\label{tab:stride30}\end{table}

\begin{table}[!htbp]\centering\small\setlength{\tabcolsep}{4pt}
\caption{Selection of $K$ at stride 30. TRE (mm) of the hyperelastic term on development cases 8 / 1}
\begin{tabular}{ccccccc}\toprule
budget (min) & $K = 8$ & $K = 14$ & $K = 23$ & $K = 40$ & $K = 56$ & $K = 80$ \\\midrule
8.5 & -- & 3.3 / 1.2 & \textbf{3.2 / 1.2} & 3.4 / 1.3 & 3.4 / 1.3 & 3.5 / 1.3 \\
4.25 & -- & 3.6 / 1.1 & \textbf{3.4 / 1.2} & 3.6 / 1.3 & 4.0 / 1.3 & 4.1 / 1.3 \\
2 & -- & 4.0 / 1.2 & \textbf{3.9 / 1.2} & 4.4 / 1.4 & 4.8 / 1.3 & 5.0 / 1.4 \\
1 & 4.8 / 1.3 & \textbf{4.5 / 1.2} & 4.7 / 1.3 & 5.4 / 1.4 & 5.7 / 1.4 & 6.4 / 1.4 \\
\bottomrule\end{tabular}
\label{tab:supp-s30k}\end{table}

\subsection{The incidence of folding}
\label{sec:fields}
\begin{figure}[!htbp]\centering
\IfFileExists{figures/folding_lattice.pdf}{\includegraphics[width=\linewidth]{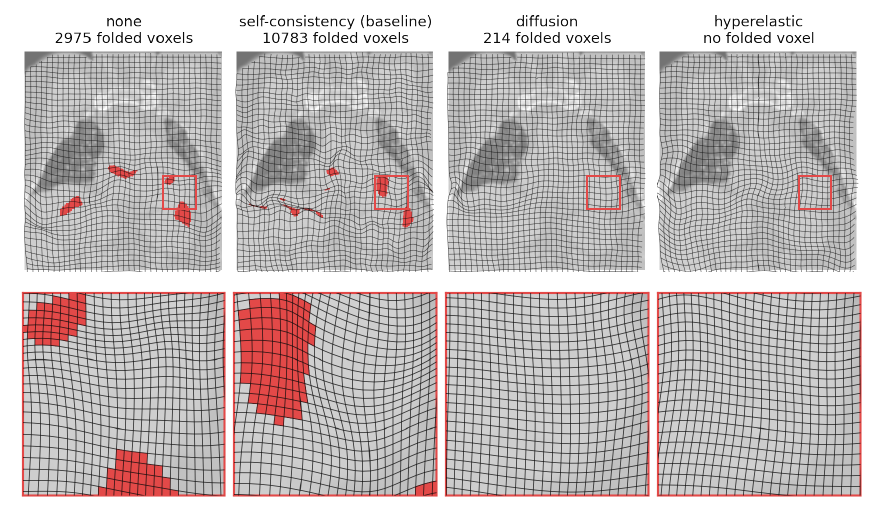}}{\tbd{figure pending}}
\caption{Folding of deployed fields. Held-out case~6 at end-exhale. A regular 2~mm lattice of the
predicted volume, mapped into the reference by each predicted field (every 4th line above, every line
in the zooms). Red indicates voxels with $\det J \le 0$. The plane and the $48 \times 48$~mm box, enlarged below,
are those with the most baseline folds, and are the same for every regularizer.}
\label{fig:lattice}
\end{figure}

Table~\ref{tab:fields} reports the folding of the predicted fields at every
held-out view of every held-out case, with $\det J$ on the full grid. The
hyperelastic term produced no folded voxel in any view, with soft-tissue $\det J$ between 0.84 and
1.17 (1st--99th percentile). Baseline fields folded in 83\% of views, over 0.43\% of the body on
average, with soft-tissue $\det J$ spanning 0.25--1.72. SVF and bending energy were also nearly
fold-free (folds in 1\% and 5\% of views), and $|\log J|$ folded in 27\% of views, though
negligibly in extent (below 0.001\% of the body). As shown in figure~\ref{fig:lattice}, on case~6 at end-exhale, the baseline field folds in a band across the lower lung and in isolated patches, where lattice lines
cross, and without a regularizer it folds in similar places.

\begin{table}[!htbp]\centering\small\setlength{\tabcolsep}{3pt}
\caption{Folding of encoder-predicted fields. Every held-out view of the held-out cases, mean over cases.}
\begin{tabular}{lccccc}\toprule
regularizer & \shortstack{fold,\\body (\%)} & \shortstack{fold,\\lung (\%)} & \shortstack{views\\folded (\%)} & \shortstack{soft-tissue $\det J$,\\p1--99} & \shortstack{lung $\det J$,\\p1--99} \\\midrule
none & 0.550 & 0.170 & 73 & 0.22--1.88 & 0.51--1.91 \\
self-consistency (baseline) & 0.425 & 0.048 & 83 & 0.25--1.72 & 0.57--1.67 \\
SVF & {\boldmath\bfseries 0.000} & {\boldmath\bfseries 0.000} & 1 & 0.55--1.58 & 0.71--1.80 \\
bending & 0.001 & {\boldmath\bfseries 0.000} & 5 & 0.74--1.24 & 0.80--1.51 \\
diffusion & 0.011 & 0.019 & 54 & 0.66--1.37 & 0.67--1.66 \\
inverse consistency & 0.003 & 0.001 & 26 & 0.66--1.36 & 0.73--1.57 \\
isotropic TV & 0.023 & 0.032 & 62 & 0.70--1.27 & 0.71--1.63 \\
$|\log J|$ & {\boldmath\bfseries 0.000} & 0.001 & 27 & 0.87--1.17 & 0.86--1.70 \\
{\boldmath\bfseries hyperelastic} & {\boldmath\bfseries 0.000} & {\boldmath\bfseries 0.000} & {\boldmath\bfseries 0} & 0.84--1.17 & 0.86--1.31 \\
\bottomrule\end{tabular}
\label{tab:fields}\end{table}

\subsection{The effect of enforcing self-consistency}
Without any regularizer (descent-only c-step), the case-8 penalty for dropping self-consistency fell
from 2.2 to 0.7 to 0.2~mm as the finest lattice was coarsened from 24 to 16 to 12 knots, and
folding fell with it (6.1, 2.1 and 0.75\% of soft tissue). Without self-consistency, 4.9\% of the
lung core folded at T50 on case 8, against 0.03\% with it, and the folds were not concentrated at
the pleura. When the solved scores were scaled by $s \in [0.85, 1.15]$, self-consistency changed by
a factor of 1.30, against 3.35 for the hyperelastic term, as expected, since for small deformations $\psi(\det J)$ grows as the fourth power of $\nabla\cdot\uu$. The induced bias in
the optimal scale was $-1.4\%$ for self-consistency against $-0.9\%$ for the hyperelastic term on
case 8, and $-3.9\%$ against $-0.7\%$ on case 1.

\section{Discussion}

In a paradigm where 2D--3D models are trained immediately prior to image-guided interventions, training time takes on special significance. In this study, we examined the performance of nine regularizers for this task. On two development cases, the hyperelastic term, paired with a descent-only basis update, reduced the time per stage-1 iteration by a third, reached the held-out accuracy of the baseline in a quarter to a half of the published budget, and produced deployed fields that did not fold. The published configuration requires nearly 10~min of training in total. With the hyperelastic term, about 3~min gave the same held-out accuracy (Figure~\ref{fig:budget}), and about 2~min when stage~1 solved on 23 projections (Table~\ref{tab:stride30}). This saving could be used to either improve the total time for the intervention or to improve accuracy. Indeed, while the baseline method did not improve on the held-out cases beyond the published budget, the hyperelastic term continued to improve, to 2.3~mm at 17~min.

In 3D--3D registration, the hyperelastic volume term was introduced to exclude folding without penalizing rigid motion or shear \cite{burger2013}. Our results suggest that this property carries over to 2D--3D
models fitted from line integrals. Penalties that act only once a voxel has folded, the
negative-Jacobian hinge \cite{mok2020} and the control-point injectivity bound \cite{choi2000}, were
no better than no regularizer, which suggests that the regularizer must shape the fields before they
fold. The stationary velocity field was nearly fold-free by construction
\cite{arsigny2006,dalca2019} but the least accurate, possibly because exponentiating the field
changes the motion that the low-rank basis can express. Isotropic TV, proposed for sliding motion
\cite{vishnevskiy2017}, did not help on the case with the most motion, and the quadratic penalties
\cite{rueckert1999,balakrishnan2019} were within 0.1~mm of the baseline. First-order inverse
consistency, which needs no image, inverse field or resampling, was the next most accurate and lost
least to self-consistency on case~6, but it folded in a quarter of deployed views.

Our experiments also suggest a rationale as to why a volume penalty can replace an image-space term. Self-consistency can be viewed as an inverse-consistency prior \cite{shao2025dreme}, but its main effect appears to be the suppression of folding by the finest control-point lattice. The accuracy lost by removing it fell from 2.2 to 0.2~mm as that lattice was coarsened, in step with the folded fraction, and the folds it prevented lay in the lung core rather than at the pleura. A volume penalty addresses this directly, on the deformation, without an inverse field or a second warp. The two terms nevertheless required different basis updates
(Table~\ref{tab:e2}). The hyperelastic term diverged under the Adam c-step and was best under the
descent-only one, whereas self-consistency showed the reverse.

Folding has direct clinical consequences. A folded field maps two points of the reference to one, so the propagated contour, mesh or accumulated dose is no longer a faithful image of the original. None of the deployed hyperelastic fields folded, whereas most deployed baseline fields did (Table~\ref{tab:fields}). Folding was nevertheless penalized rather than excluded. The barrier was continued by a finite quadratic, evaluated on a subsampled grid and not imposed on the encoder, so the absence of folding is a measured property of these cases rather than a strict guarantee.

This study has several limitations. The ventilated arcs were simulated from
ten-phase 4DCT, without scatter, hysteresis or motion between phases, and no measured patient arc was
used. However, this arises as a natural consequence of our evaluation paradigm in which 4D data are used to supply ground-truth that is not otherwise possible. Additionally, weights were selected on two cases, so differences of 0.1--0.3~mm between regularizers should not be overinterpreted. Finally, on case 6, the held-out case with the most motion, every geometric penalty was less accurate than self-consistency. We attribute this to the motion model rather than the regularizer as a smooth B-spline basis cannot express sliding of the lung against the chest wall. Exploring alternative bases will form a major direction for future work.

\section{Conclusions}
Patient-specific motion models can recover 3D anatomy from a single x-ray projection, but when the model is trained immediately prior to an intervention, training time becomes crucial. Here we showed that the regularization term can be the largest contributor to training time and that hyperelasticity offers an efficient penalty acting on the deformation itself. Replacing self consistency with the hyperelastic regularizer matched the held-out accuracy of the baseline method with 3--5~minutes of training and removed folding from the deployed fields. This brings on-the-day patient-specific motion models closer to routine use in image-guided interventions.

\section*{Acknowledgements}
This work was funded by the NSW Government through the Cancer Institute NSW and by the Australian Government by the National Health and Medical Research Council.

\section*{Code and data availability}
The code will be available at \url{https://github.com/Image-X-Institute/Voxelmap/tree/hyper}. The DIR-Lab 4DCT data and landmarks are publicly available from the DIR-Lab website: \url{https://med.emory.edu/departments/radiation-oncology/research-laboratories/deformable-image-registration/downloads-and-reference-data/4dct.html}.

\section*{Competing interests}
The author has filed a patent relating to the method disclosed in this paper.

\end{document}